\documentclass{svjour3}                     
\smartqed  

\usepackage{graphicx}
\usepackage{mathptmx}
\usepackage{latexsym}

\begin{document}

\title{On the growth of perturbations in interacting
dark energy and dark matter fluids. }

\author{N.A. Koshelev  }

\institute{ Ulyanovsk State University, Leo Tolstoy str 42, 432970, Russia \\
              \email{koshna71@inbox.ru}  }

\date{Received: date / Accepted: date}

\maketitle

\begin{abstract}
The covariant generalizations of the background dark sector
coupling suggested in G. Mangano, G. Miele and V. Pettorino, Mod.
Phys. Lett.  \textbf{A 18}, 831 (2003) are considered. The
evolution of perturbations is studied with detailed attention to
interaction rate that is proportional to the product of dark
matter and dark energy densities. It is shown that some classes of
models with coupling of this type do not suffer from early time
instabilities in strong coupling regime. 
\end{abstract}

\section{Introduction}
\label{intro} Recent observations have shown that the Universe is
spatially flat and presently accelerating. Most attempts to
explain this acceleration involve the introduction of dark energy,
as a source in the Einstein field equations. In addition, the
important contribution to the total density is the dark matter,
whose existence is inferred indirectly by observing its
gravitational influence on normal matter, such as stars, gas and
dust. Standard $ \Lambda CDM $ model, in which dark energy is
considered as a small positive cosmological constant and dark
matter treated as gas of cold non-baryonic particles, provides a
very good fit to the supernovae data \cite{Astier} as well as CMB
measurements \cite{WMAP5} and observations of large scale
structure \cite{SDSS} , but the small and fine-tuned value of the
cosmological constant cannot be explained within current particle
physics \cite{Weinberg}. As a result a lot of other more expected
from theoretical point of view cosmological models have been
proposed to giving a dynamical origin to dark energy.

Models with non-minimal coupling of dark matter and dark energy
have called attention in the last decade. The nature of both dark
energy and dark matter are still unknown, and in principle, the
additional interaction between them is possible. Moreover, the
coupling between matter and quintessence is motivated by high
energy particle physics considerations (see \cite{Hoffman_rev} for
a set of references). There are a large number of works in where
the interaction is seen in models with quintessence
\cite{Hoffman_rev,Wetterich,Amendola,Holden_Wands,Farrar_Peebles,Ziaeepour,Koivisto,08011105}
field. Also available the phenomenological approach, in which the
interaction is introduced into the conservation equations for dark
matter ($c$) and dark energy ($x$), considered as perfect fluids.
The background energy exchange in the dark sector can be
represented by
\begin{equation}
\bar{\rho}_c'=-3 {\cal H}\bar{\rho}_c+a\bar{Q},
\end{equation}
\begin{equation}
\bar{\rho}_x'=-3{\cal H}(1+w_x)\bar{\rho}_x-a\bar{Q},
\end{equation}
where the prime denotes derivative with respect to conformal time
$\tau$, the bars mark the background quantities, $a$ is the scale
factor, ${\cal H}=a'/a$, $w_x=\bar{P}_x/\bar{\rho}_x$ and the
coupling $\bar{Q}$ is some function of background variables.

In the simplest models the quantity $ \bar{Q} $ is a linear
combination of the dark sector densities
\begin{equation}
\label{coupl_backgr}\bar{Q}=A_c\bar{\rho}_c+A_x\bar{\rho}_x,
\end{equation}
where $A_I=3\alpha_I H$ or $A_I=\Gamma_I$. Here $ \alpha_I $ are
the dimensionless constants and $\Gamma_I$ are constant
interaction rates \cite{CMU}. There are a large number of works in
which for coupling (\ref{coupl_backgr}) was considered the
background \cite{CMU,Pavon2003,QCJRW} and the perturbations
evolution
\cite{VMM_0804,08073471,09011611,09013272,09050492,VMM_0907}.

In the paper \cite{0212518} in the context of the stimulated decay
of dark energy into dark matter had been proposed the interaction
of the form
\begin{equation}
\label{coupl_backgr2} \bar{Q} =\gamma \bar{\rho}_c ^\alpha
\bar{\rho}_x^\beta,
\end{equation}
where $\alpha$, $\beta$ are the dimensionless constants, and
$\gamma$ is a constant parameter with dimension
$[$density${}^{1-\alpha-\beta}$$\times$time${}^{-1}]$. This
coupling provide a mechanism to alleviate the cosmic coincidence
problem \cite{coincidence_original},  \cite{coincidence}, namely
why the dark energy density nearly coincides with the dark matter
density presently. The models with interaction
(\ref{coupl_backgr2}) also have been studied in the framework of
holographic dark energy \cite{09011215}. By assuming that $ \alpha
= 1, \beta = 0 $ or $\alpha = 0, \beta = 1 $ one obtains the
particular cases of coupling (\ref{coupl_backgr}).

Although the basic properties of background solutions for the
interaction (\ref{coupl_backgr2}) are previously considered, the
perturbation evolution in such scenarios is not examined except
the mentioned above special cases. In this paper we shall cover
the perturbation equations and investigate the solutions of them
with detailed attention to the most physically reasonable choice
of parameters $ \alpha = \beta = 1 $.

\section{Background.}
\label{sec:1} We consider the spatially flat Universe filled with
radiation $(r)$, massless neutrino $(\nu)$, baryonic matter $(b)$,
cold dark matter $(c)$ and dark energy $(x)$. We choose the linear
parametrization of the dark energy equation of state
\cite{parametrization}
\begin{equation}
\label{eos_param}w_x (a) = w_0+w_1(1-a)
\end{equation}
with constants $w_0>-1$ and $0<w_1<-w_0$, so the dark energy is
subdominant with respect to the radiation at early times and the
phantom divide line $w_x=-1$ crossing does not occur. Since
$w_x'/{\cal H}=-aw_1$, in the radiation dominated era the dark
energy equation of state parameter $w_x$ is well approximated by a
constant $w_0+w_1$.

Background dynamics in the presence of coupling
(\ref{coupl_backgr2}) is completely described by Friedmann's
equation
\begin{equation}
\label{Friedmann}{\cal H}^2=\frac{8\pi G}{3}a^2\bar{\rho}
\end{equation}
and continuity equations
\begin{equation}
\label{continuity_rad}\bar{\rho}_r'=-4{\cal H}\bar{\rho} _r, ~~~~
\bar{\rho}_\nu'=-4{\cal H} \bar{\rho}_\nu,
\end{equation}
\begin{equation}
\bar{\rho}_b'=-3{\cal H}\bar{\rho}_b,
\end{equation}
\begin{equation}
\label{continuity_dm}\bar{\rho}_c'=-3{\cal
H}\bar{\rho}_c-a\gamma\bar{\rho}_c^\alpha \bar{\rho}_x^\beta,
\end{equation}
\begin{equation}
\label{continuity_de}\bar{\rho}_x'=-3{\cal H}(1+w_x)\bar{\rho}_x+
a\gamma \bar{\rho}_c^\alpha \bar{\rho}_x^\beta .
\end{equation}

To determines the characteristic strength of the dark sector
coupling it is convenient to introduce the dimensionless parameter
$\lambda$ defined by \cite{09011215}
\begin{equation}
\lambda = \gamma\rho_{0cr}^{\alpha+\beta-1}H_0^{-1}
\end{equation}
where $\rho_{0cr}=3H_0^2 /(8\pi G) $ is the current critical
density.

System of equations (\ref{Friedmann})-(\ref{continuity_de}) can be
numerically solved, as was done in \cite{0212518} for $ \alpha = 1
$ and the dark energy equation of state $ w_x =- 1 $. However, the
equations (\ref{continuity_dm}) and (\ref{continuity_de})
immediately suggest several analytical conclusions.

When $ \gamma < 0$, the dark energy density drops with the scale
factor $a(\tau)$ not slower than $a^{-3(1+w_0)}$ and the dark
matter density can increase under  $ \left|a\gamma
\bar{\rho}_c^{\alpha -1} \bar{\rho}_x^\beta/{\cal H}\right|>3$
condition. These features offer a simple way out to the cosmic
coincidence problem, since in this models the present situation of
$\bar{\rho}_x \sim \bar{\rho}_c$ have occurred many times in the
past \cite{0212518}.

The class of models with positive $\gamma$ also describe a
interesting scenarios, as it allow to consider the dark energy
grew out due the dark matter decay. However, the interaction of
this type potentially leads to unphysical results, because at
early times the dark energy density could be driven to negative
value in some models with decaying dark matter \cite{VMM_0804}.

\begin{figure}
\includegraphics[width=0.48\textwidth]{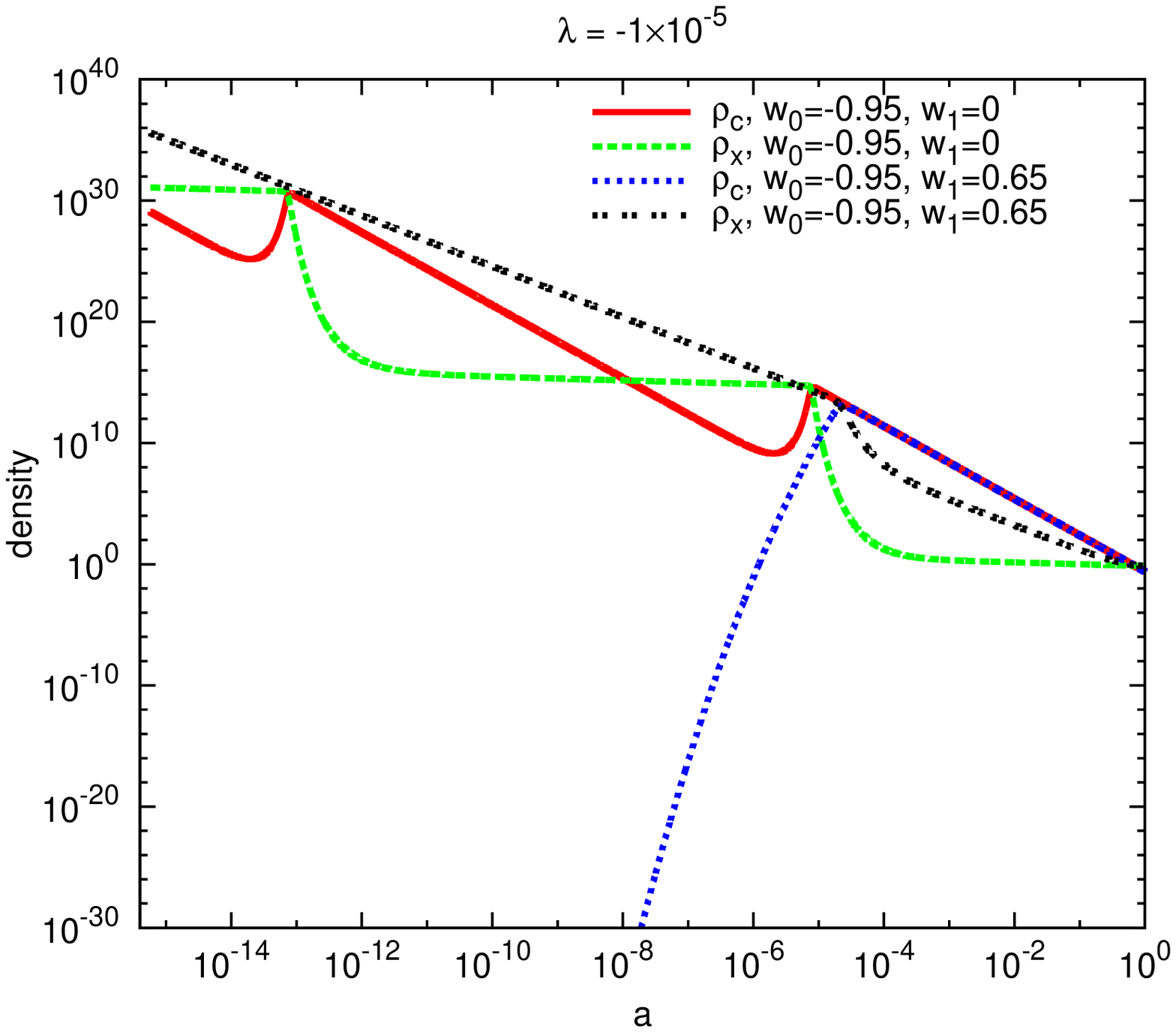}
\includegraphics[width=0.48\textwidth]{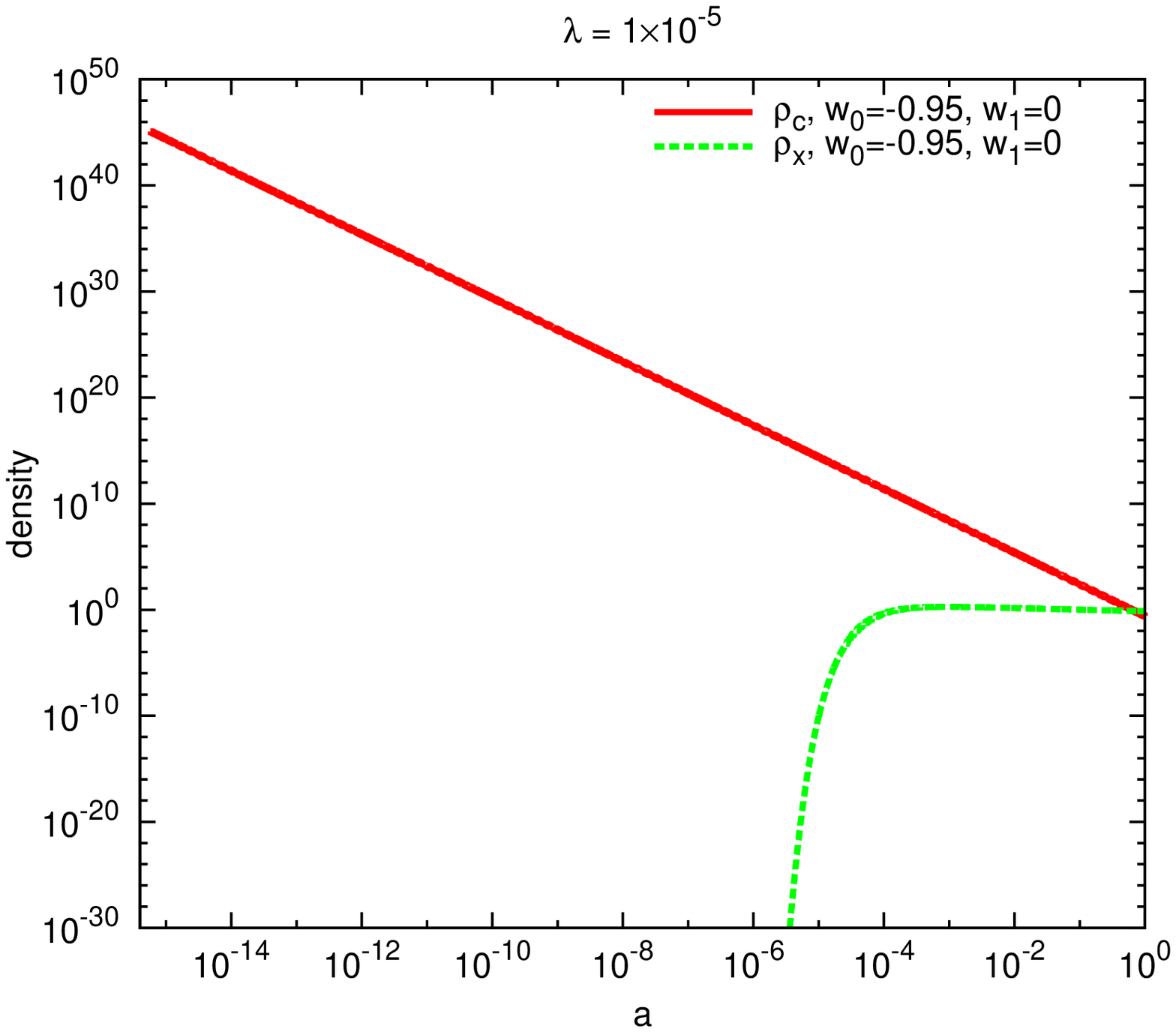}
\caption{Background energy densities in units of $\rho_{0cr}$ for
$\alpha=\beta=1$, $H_0=70$ km s${}^{-1}$Mpc${}^{-1}$,
$\Omega_x=0.70$, $\Omega_c=0.2538$, $\Omega_b=0.0462$. Examples of
dark sector densities evolution at negative and positive coupling
are shown on left and right panels, respectively.  } \label{fig:1}
\end{figure}

Figure 1 shows the time evolution of the dark matter and dark
energy densities at values parameters $ \alpha = \beta = 1 $. The
pattern of the solutions varies depending on the model parameters.
When $w_x< -1/3$ at high redshifts and $ \gamma <0$,
one can obtain a sequence of regimes of the weak \\
$max(\left|a\gamma \bar{\rho}_c /{\cal H}\right|, \left|a\gamma
\bar{\rho}_x /{\cal H}\right|)\ll 1$ and strong $max(\left|a\gamma
\bar{\rho}_c /{\cal H}\right|, \left|a\gamma \bar{\rho}_x /{\cal
H}\right|)\geq 1$ dark sector coupling. At the stage of the weak
interaction under the condition $w_x<-1/3$ the dark energy density
decays more slowly than $a^{-2}(\tau)$, and hence in the radiation
dominated era the inequality $\left|a\gamma \bar{\rho}_x/{\cal
H}\right|>3$ will be reached at some time, from which the density
of dark matter increases. The increased dark matter density
increases the dark energy decay rate, the condition $
\left|a\gamma \bar{\rho}_x/{\cal H}\right|\ll 1$ recovers and dark
energy has no longer significant impact on the evolution of dark
matter. Moreover, since the dark matter density is now reduced
proportional to the inverse third degree of the scale factor,
there comes a point when the influence of dark matter to dark
energy also becomes negligible, i.e. a next stage of a weak
interaction begins. For $-1/3 <w_x$ however, if the interaction is
weak initially, it will remain so through the subsequent
evolution. This means that the stage of strong interaction in such
models is only one. Accordingly, in such models the density of
dark matter in the radiation dominated era is extremely small.

The right panel of Figure \ref{fig:1} depicts the time evolution
of the density of dark matter and dark energy at values parameters
$ \alpha = \beta = 1 $ and positive $\gamma$. In this case, the
density of dark energy is always positive.

\section{Model and perturbed equations.}
\label{sec:2}

Let us consider a spatially flat FLRW Universe with small scalar
type perturbations. The line element may be written as
\begin{equation} ds^2 =
a^2(\tau)\left\{-(1+2\phi) d\tau^2 +2B_{,i}d\tau dx^i+[(1 - 2\psi)
\delta _{ij} +2E_{,ij}]dx^idx^j \right\}.
\end{equation}

The energy-momentum tensor of a perfect $A$-fluid is given by
\begin{equation}
T_{A\nu} ^{\mu}  = \left(\rho_A + P_A \right)u_A^{\mu} u_{A\nu} +
P_A\delta _{\nu}^{\mu}  ,
\end{equation}
where $\rho_A =\bar{\rho}_{A}(1+\delta_{A}) $ is density,
$P_A=\bar{P}_A+\delta P_A$ is pressure, and the fluid
four-velocity $u_A^{\mu}=dx_A^\mu/ds $ at linear order of the
perturbations is
\begin{equation}
u_A^{\mu} = \frac{1}{a} \left[ \left(1-\phi\right), ~v_{A}^{,i}
\right] , ~~~~~~ u_{\nu} = a \left[ -(1+\phi),~v_{A,i}+B_{,i}
\right].
\end{equation}

We assume that the anisotropic stress of dark energy and dark
matter vanishes, so the dark species can be treated as perfect
fluids. In the general case of coupling fluids divergence of the
energy-momentum tensor of each fluid yields
\begin{equation}
\label{div_eq}T^{\mu\nu}_{A;\nu}=Q^{\mu}_{A},
\end{equation}
where 4-vectors $Q^{\mu}_{A}$ are restricted by the constraint
\cite{Kodama_Sasaki}
\begin{equation}
\sum_A Q_A^\mu =0,
\end{equation}
which follow from the conservation law of the total
energy-momentum tensor. For convenience one can decompose these
4-vectors into two parts
\begin{equation}
Q_A^\mu =Q_A u^\mu +F_A^\mu, ~~~~ Q_A=\bar{Q}_A+ \delta Q_A,
~~~~u_\mu F_A^\mu =0.
\end{equation}
Here $u^\mu$ is the total four-velocity, $Q_A$ is the energy
density transfer rate and $F_A^\mu$ is the momentum density
transfer rate of $A$-fluid in the total matter gauge. By
definition, $F_A^\mu =a^{-1}(0,f_A^{~,i})$, where $f_A$ is a
momentum transfer potential \cite{VMM_0804}.

The energy exchange in the background does not determine the
covariant form of energy exchange. Instead, an energy exchange
four-vectors $Q^{\mu}_{A}$ must be specified. The simplest
scalars, which can be used to construct quantities $Q^{\mu}_{c}$,
$Q^{\mu}_{x}$ are
\begin{equation}
T^\sigma_{c~\!\sigma} =\rho_c  , ~~~~~~ T^\sigma_{x~\!\sigma}
=\rho_x-3p_x , ~~~~~~T^\sigma_{c~\!\nu} T^\nu_{x~\!\sigma} =\rho_c
\rho_x
\end{equation}
where the last relation holds at linear order. Accordingly, the
direct generalization (\ref{coupl_backgr2}) with
\begin{equation}
Q_c=-Q_x=-\gamma \rho_c^\alpha\rho_x^\beta,
\end{equation}
is covariant at the first order.

We are interested first of all in models with decaying dark matter
or with decaying dark energy. The most natural way is to choose a
vector $F_A^\mu$ in the form
\begin{equation}
F_c^\mu=-F_x^\mu=\frac{1}{a}\left(0,(\gamma
\rho_c^\alpha\rho_x^\beta(v-v_c))^{,i}\right)
\end{equation}
or
\begin{equation}
F_c^\mu=-F_x^\mu=\frac{1}{a}\left(0,(\gamma
\rho_c^\alpha\rho_x^\beta(v-v_x))^{,i}\right).
\end{equation}
The momentum transfer vanishes in the dark matter rest frame or in
the dark energy rest frame respectively. A few more general form
of coupling can be written as
\begin{equation}
\label{coupl_1} Q^\mu_x=-Q^\mu_c = \rho_c^\alpha\rho_x^\beta
(\gamma_c u^\mu_c + \gamma_x u^\mu_x),
\end{equation}
where $\gamma_c$ and $\gamma_x$ are arbitrary constants.

The continuity equations for coupled perfect fluids can be
obtained by linearization of equations (\ref{div_eq}). As a
result, one can write \cite{VMM_0804}
\begin{equation}
\delta\rho_{A}' + 3{\cal H}\left(\delta\rho_{A}+\delta
P_{A}\right) -  3\left( \bar{\rho}_{A} + \bar{P}_{A}\right) \psi'
-k^2 \left( \bar{\rho}_{A} + \bar{P}_{A}\right) \left(v_A+E'
\right) = a \bar{Q}_A\phi +a\delta Q_A,
\end{equation}
\begin{equation}
\left[(\bar{\rho}_A\!+\bar{P}_A) (v_A+B) \right]' + 4{\cal
H}(\bar{\rho}_A\!+ \bar{P}_A)(v_A+B)+ (\bar{\rho}_A\!+
\bar{P}_A)\phi +\delta P_A = a \bar{Q}_A(v+B) +af_A ,
\end{equation}
where for coupling (\ref{coupl_1})
\begin{equation}
\delta Q_c=-\delta Q_x=-(\gamma_c+\gamma_x) \bar{\rho}_c^\alpha
\bar{\rho}_x ^\beta (\alpha \delta_c+\beta\delta_x),
\end{equation}
\begin{equation}
f_c=- f_x= \bar{\rho}_c^\alpha\bar{\rho}_x^\beta \left(
\gamma_c(v-v_c) +\gamma_x(v-v_x)\right).
\end{equation}

Pressure perturbations in the general case can be expressed in
terms of the density perturbations and the velocity potentials
\begin{equation}
\delta P_A  =c^2_{sA}\delta \rho_A + (c^2_{sA}-c^2_{sA(ad)})
\left(3{\cal H} (1+w_A)\rho_A-a \bar{Q}_{A}\right)
\frac{\theta_A}{k^2}.
\end{equation}
where $\theta_A =-k^2(B+v_A)$, $c_{As(ad)} ^2 = P_A'/\rho_A'$ is
the adiabatic sound speed and the rest frame sound speed $s_{sA}$
defined by
\begin{equation}
c_{sA} ^2 =\left.\frac{\delta P_A}{\delta \rho_A}
\right|_{A-fluid~rest~frame}.
\end{equation}

We will work in synchronous orthogonal gauge $\phi = 0$, $B = 0$
in the Fourier space, using the notation of
\cite{Ma_Bertschinger}:
\begin{equation}
\psi_{\mathbf{k}} =\eta, ~~~~~~ k^2E_{\mathbf{k}}=-h/2 - 3\eta .
\end{equation}

Now the conservation equations (\ref{div_eq}) for the dark energy
and dark matter in the synchronous gauge takes the form
\begin{eqnarray}
\label{basic_eq1}\delta_x '+ 3{\cal H}(c^2_{sx}-w_x) \delta_x +
\left( 1 + w_x \right)\theta_x + 3{\cal H}\left[3{\cal H} (1+w_x)
(c^2_{sx}-w_x) +w_x' \right]\frac{\theta_x}{k^2}   \nonumber \\
+\frac{1 + w_x}{2} h'=(\gamma_c+\gamma_x)a\bar{\rho}_c^{\alpha}
\bar{\rho}_x^{\beta -1}\left[\alpha \delta_c +(\beta-1)\delta_x +
3{\cal H} ( c^2_{sx}-w_x) \frac{ \theta_x}{k^2}\right] ,
\end{eqnarray}
\begin{equation}
\label{basic_eq2}\delta_c ' + \frac{1}{2}h' +\theta_c =(\gamma_c+
\gamma_x) a\bar{\rho}_c ^{\alpha -1} \bar{\rho}_x^{\beta} \left[
(1-\alpha) \delta_c -\beta\delta_x \right],
\end{equation}
\begin{equation}
\label{basic_eq3}\theta'_x + {\cal H}\left(1 - 3c^2_{sx}\right)
\theta_x -\frac{c_{sx}^2k^2}{1+w_x } \delta_x = \frac{a \bar{\rho}
_c ^\alpha\bar{\rho}_x^{\beta -1}}{ 1+w_x} \left (\gamma_c
(\theta_c-\theta_x)-(\gamma_c+\gamma_x)c^2_{sx} \theta_x\right) ,
\end{equation}
\begin{equation}
\label{basic_eq4}\theta_c'+{\cal H}\theta_c = \gamma_x a
\bar{\rho}_c^{\alpha -1}\bar{\rho}_x^{\beta}[\theta_c-\theta_x].
\end{equation}
Note that in this phenomenological approach the sound speed
$c_{sx}$ is needed to be fixed by hand \cite{Gordon_Hu}. In the
case of quintessence dark energy one have to set $c_{sx}=1$, and
we adopt this value in the following.

The perturbed Einstein equations are well known, and can be found
in \cite{Ma_Bertschinger}. We reproduce here only one of them
\begin{equation}
\label{Einstein_pert}h'' + {\cal H}h'= -8\pi G ^2a^2 (\delta \rho
+ 3\delta P ).
\end{equation}
Equations (\ref{basic_eq1}) - (\ref{basic_eq4}) provide a set of
coupled equations covering the dark sector density evolution.

\section{Large scale perturbations.}
\label{sec:3} The coupling terms appearing in the dark energy
pressure perturbations may lead to the early time instabilities,
as was first pointed out in \cite{VMM_0804}. The similar
phenomenon is well known in inflationary multi-fields models,
where on large scales entropy perturbations can source adiabatic
ones \cite{GWBM}. To study this phenomenon here we write the
second order differential equations for the dark energy density
perturbations. This approach \cite{09011611} allows to identify
areas of possible instabilities before solving the perturbation
equations.

The perturbed fluid equations (\ref{basic_eq1}) -
(\ref{basic_eq3}) and equation (\ref{Einstein_pert}) can be
combined to
\begin{equation}
\label{de_pert} \delta_x ''+ A_x{\cal H} \delta _x' + B_x{\cal
H}^2 \delta _x =C_x {\cal H}^2,
\end{equation}
where on large scales ($k\ll {\cal H}$)
\begin{equation}
A_x= 1-3w_x -\frac{2{\cal H}'}{{\cal H} ^2} + \left(\frac{\gamma_c
+ \gamma_{+}}{1+w_x} - (\beta-1) \gamma _{+}\right)
\frac{a\bar{\rho}_c ^{\alpha} \bar{\rho}_x^{\beta -1}}{{\cal H}} -
f,
\end{equation}
\begin{eqnarray}
B_x = 3(1- w_x)\left(1 - \frac{{\cal H}'}{{\cal H}^2}+ \frac{
\gamma_c a\bar{\rho}_c ^{\alpha} \bar{\rho}_x^{\beta -1}}{(1
+w_x){\cal H}}- f\right)-
\frac{3w_xw_x'}{(1+w_x){\cal H} }  \nonumber\\
+ (\beta-1) \gamma_{+} \frac{a\bar{\rho}_c ^{\alpha}
\bar{\rho}_x^{\beta -1}}{{\cal H}} \left(2+ \frac{{\cal H}'}{{\cal
H}^2} -\frac{\gamma_c + \gamma_{+} }{1+w_x} \frac{a\bar{\rho}_c
^{\alpha}\bar{\rho}_x^{\beta -1}}{{\cal H}}+f \right)
\nonumber\\
-(\beta-1)\gamma_{+}\frac{1}{{\cal H}}\left(\frac{a \bar{\rho}_c
^{\alpha} \bar{\rho}_x^{\beta -1}}{{\cal H}}\right)'+\gamma_{+}^2
\alpha\beta \frac{a^2\bar{\rho}_c^{2\alpha -1}
\bar{\rho}_x^{2\beta-1} }{{\cal H}^2},
\end{eqnarray}
\begin{eqnarray}
C_x = \frac{1}{2} ( 1 + w_x)\left(  3  + \frac{2{\cal H}'}{ {\cal
H}^2} -\frac{\gamma_c +\gamma _{+} (\alpha + 1)}{1 + w_x}
\frac{a\bar{\rho}_c ^{\alpha} \bar{\rho}_x^{\beta -1}}{{\cal H}} +
f\right)\frac{h'}{{\cal H}}- \frac{w_x'h'}{2{\cal H}^2} \nonumber \\
- 3{\cal H}\left( 3-3w_x+\frac{w_x'}{ (1+w_x){\cal H}}  -
\frac{1-w_x}{ 1+w_x}\frac{\gamma_{+}a \bar{\rho}_c ^{\alpha}
\bar{\rho}_x ^{\beta -1}}{{\cal H}} \right) \frac{\gamma_c a
\bar{\rho}_c^\alpha \bar{\rho}_x^{\beta
-1}}{{\cal H}}\frac{\theta_c}{k^2} \nonumber \\
+\frac{\gamma_{+}}{{\cal H}}\left(\frac{a \bar{\rho}_c ^{\alpha}
\bar{\rho}_x^{\beta -1}}{{\cal H}}\right)'\alpha\delta_c -\left( 2
+ \frac{{\cal H}'}{ {\cal H}^2} -\frac{ \gamma _c + \gamma_{+}
}{1+w_x} \frac{a\bar{\rho} _c^ {\alpha} \bar{\rho}_x^{\beta
-1}}{{\cal H}} +f\right) \frac{\gamma_{+} a \bar{\rho}_c ^{\alpha}
\bar{\rho} _x^{\beta -1}}{{\cal
H}}\alpha\delta_c \nonumber \\
+ (1-\alpha)  \frac{\gamma_{+}^2 a^2 \bar{\rho}_c ^{2\alpha -1}
\bar{\rho} _x^{2\beta -1}}{{\cal H}^2}\alpha\delta_c+\frac{3}{2}
\left( 1 + w_x \right) \left( \delta + 3\frac{\delta P}{\rho}
\right),
\end{eqnarray}

and
\begin{equation}
\label{f_value} f=\frac{1}{{\cal H}}\left[\ln \left|
(1-w_x)\left(3+3w_x- \frac{\gamma_{+} a\bar{\rho}_c ^{\alpha}
\bar{\rho}_x^{\beta -1}}{{\cal H}}\right)  + \frac{w_x'}{{\cal
H}}\right| \right]', ~~ \gamma_{+}=\gamma_c+\gamma_x.
\end{equation}

In case of minimal coupling and a constant dark energy equation of
state the quantity $f$ is zero at all times. When $\gamma_c=0$ or
$\gamma_x=0$, terms with $\theta_c$ can be ignored, since in the
latter case one can work in particular synchronous orthogonal
gauge in which the dark matter fluid has a vanishing velocity.

The equation (\ref{de_pert}) is written in slightly different form
then was discussed in the Ref. \cite{09011611}. In particular, it
involves $h'$ instead of the time derivative of dark matter
density perturbations. This difference can be valuable, since both
quantities $h'$, $\delta_c'$ are related by equation
(\ref{basic_eq2}) and, in the general case, may implicitly depend
on the dark energy perturbations. Used here notation is convenient to
study of generations of non-adiabatic perturbations in the
radiation dominated era.

At first approximation, the source terms can be calculated using
usual adiabatic mode solutions. Since the dark species are
subdominant in early Universe, the corresponding contributions to
$h'$ and total $\delta$, $\delta P$ are negligible. Actually, this
is the assumption about the initial conditions, but it is
certainly satisfied if the initial adiabatic conditions are
imposed at the stage of weak coupling. This assumption breaks down
if the dark energy perturbations have increased dramatically. It
means that the right hand side of the equation (\ref{de_pert}) can
be treated at early times and on initial stages of dark energy
inhomogeneities growth as an external force that is independent of
the dark energy perturbations. In this approach, the negative sign
of $A_x$, $B_x$ or both of them indicate on the existence of large
scale instabilities due a anti-damping force or exponential growth
of intrinsic dark energy perturbations. The nature of instability
can be revealed by considering the time evolution of the
gauge-invariant curvature perturbation on uniform density
hypersurfaces
\begin{equation}
\zeta \equiv -\psi -{\cal H}\frac{\delta\rho}{\bar{\rho}}
\end{equation}
that is conserved on large scales for adiabatic perturbations.

We carry out the detailed analysis of perturbations for two cases:
$ \gamma_c = \gamma, \gamma_x = 0 $ and $ \gamma_c = 0, \gamma_x =
\gamma $ with $ \alpha = \beta = 1 $, i.e. for
\begin{equation}
\label{coupl_2} Q^\mu_x=-Q^\mu_c =\gamma \rho_c\rho_x(b u^\mu_c
+(1-b)u^\mu_x)
\end{equation}
where $ b = 1 $ for the first case and $ b = 0 $ for the second.

For coupling above
\begin{equation}
A_x= 1-3w_x - \frac{2{\cal H}'}{{\cal H} ^2} + \frac{b + 1}{
1+w_x} \frac{\gamma a\bar{\rho}_c }{{\cal H}}-f,
\end{equation}
\begin{equation}
B_x= 3(1- w_x)\left(1 - \frac{{\cal H}'}{{\cal H}^2}+ b
\frac{\gamma a\bar{\rho}_c}{(1 +w_x){\cal H}}- f\right)-
\frac{3w_x w_x'}{(1+w_x){\cal H} } +\gamma^2 \frac{a^2\bar{\rho}_c
\bar{\rho}_x }{{\cal H}^2} .
\end{equation}
Using equations (\ref{eos_param}), (\ref{continuity_dm}),
(\ref{continuity_de}) and (\ref{f_value}), the numerical values of
these coefficients can be calculated directly from the background
solutions.

When $ \gamma > 0 $, in the radiation dominated era one can assume
$\frac{\gamma a \bar{\rho}_x}{{\cal H}}\ll 1$, $\frac{\gamma a
\bar{\rho}_c}{{\cal H}}\gg 1$ (see the right panel of Figure
\ref{fig:1}). In this limiting case the dark energy has no effect
on the evolution of other fractions and their perturbations. By
applying equation (\ref{continuity_dm}), quantity $f$ can be
approximated as
\begin{equation}
f\approx -2 - \frac{a\gamma \bar{\rho}_x}{{\cal H}}-\frac{{\cal
H}'}{{\cal H}^2}.
\end{equation}

Keeping only the dominant terms, the equation (\ref{de_pert})
reduced to
\begin{equation}
\label{eq_simpl1}\delta_x ''\! +\! {\cal H}\!  \frac{b + 1}{1 \!+
\!w_x} \frac{\gamma a\bar{\rho}_c}{{\cal H}} \delta_x ' + \! 3
{\cal H}^2 (1\!-\! w_x)\left(\!3+\!\frac{b }{1\!+\!w_x}
\frac{\gamma a\bar{\rho} _c}{{\cal H} } \!\right)\! \delta_x \!=\!
{\cal H}^2 \!\frac{ b + 1}{1+w_x}\left(\!\frac{\gamma a
\bar{\rho}_c }{{\cal H}} \!\right)^2\!\delta_c.
\end{equation}
Coefficients $A_x$ and $B_x$ are positive and catastrophic growth
perturbations do not occurs. In order of magnitude, equation
(\ref{eq_simpl1}) implies the estimation
\begin{equation}
\delta_x \sim \frac{\gamma a\bar{\rho}_c }{{\cal H}} \delta_c .
\end{equation}
For example, the "standard" adiabatic condition $
\delta\rho_c/\bar{\rho}_c'= \delta\rho_x/\bar{\rho}_x' $ taking
into account the background equations
(\ref{continuity_dm}),(\ref{continuity_de}) in the radiation
dominated era yields
\begin{equation}
\delta_x= - \frac{\gamma a\bar{\rho}_c}{3 {\cal H}}\delta_c
\propto \tau
\end{equation}
as for the adiabatic mode in the synchronous gauge
$\delta_c\propto\tau^2$.

When $\gamma < 0$, the coefficients $A_x$, $B_x$ can take large
negative values in regime of very strong coupling if $|\gamma
a\bar{\rho}_c /{\cal H}|\gg 1$. In scenarios with $b=1$ and
$w_x(a) <-1/3$  at early times they are becoming both negative
together under this condition . Hence such models suffer from the
fast growth of non-adiabatic perturbations in early Universe.
However, constraints on coupling with $b=0$ can be weakened,
because in this case in some range of parameters $w_0$, $w_1$ the
coefficient $B_x$ remains positive and increases in the strong
coupling regime.

\begin{figure}
\includegraphics[width=0.48\textwidth]{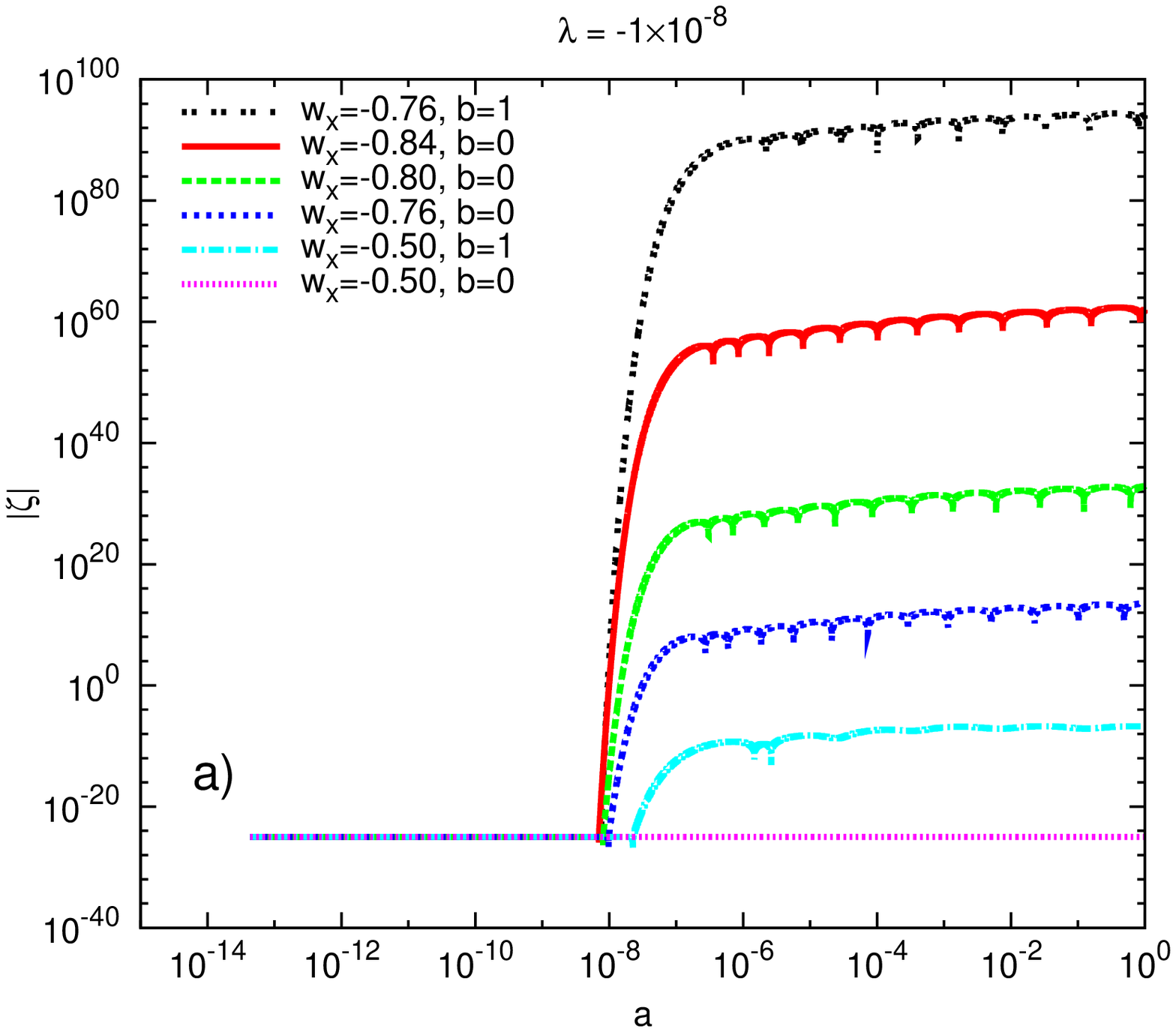}
\includegraphics[width=0.48\textwidth]{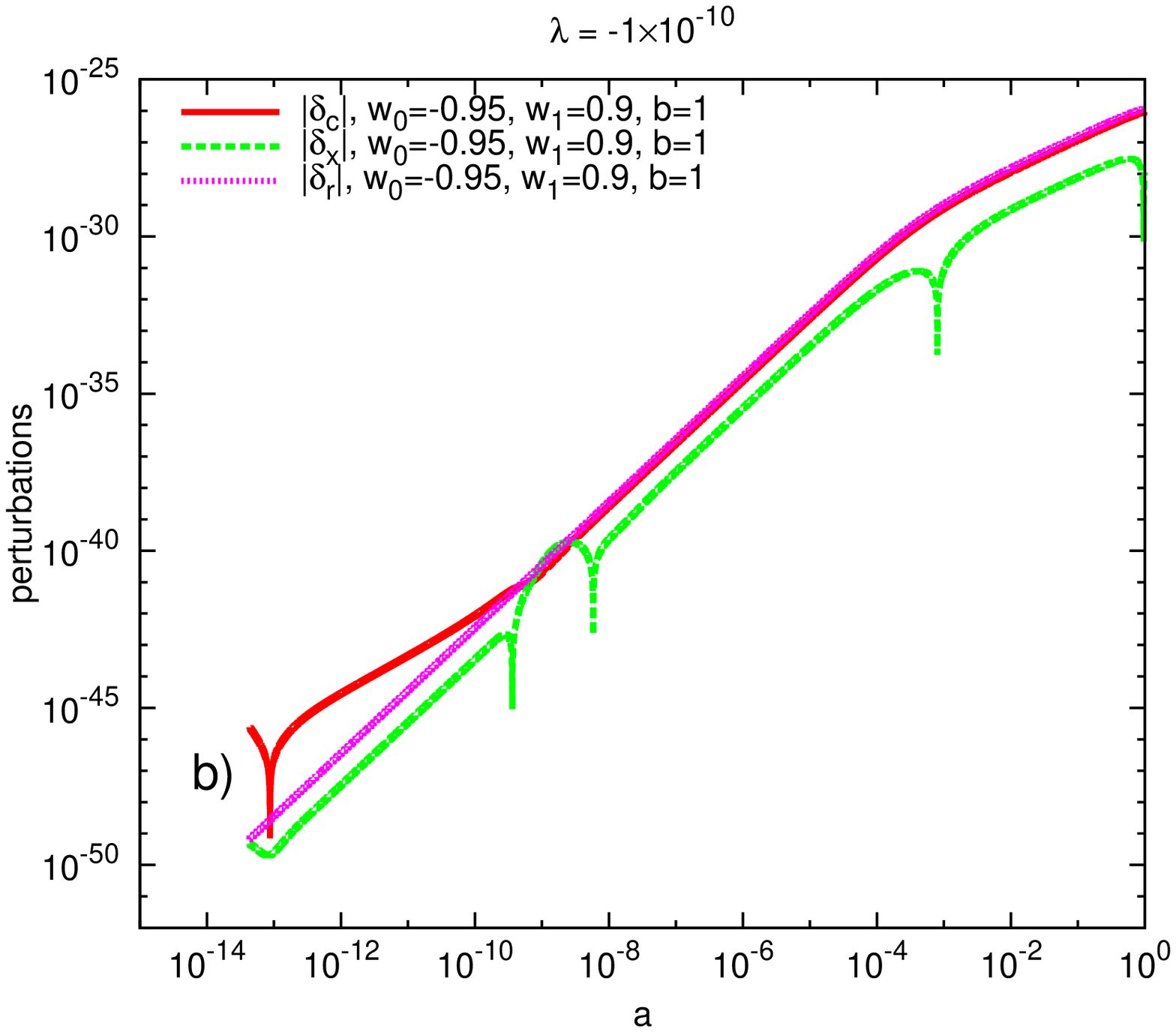}
\caption{a) The curvature perturbation $\zeta$ on super-Hubble
scale in models with coupling (\ref{coupl_2}) and constant dark
energy equation of state $w_x<-1/3$. Growth of the non-adiabatic
perturbations begins at the stage of the strong interaction. b)The
density perturbation evolution in model with $w_x>-1/3$ in
radiation dominated era. Initial dark sector conditions are set at
$\tau_{in}=2\times 10^{-8}$ Mpc by $ \delta\rho_c/\bar{\rho}_c'=
\delta\rho_x/\bar{\rho}_x'= \delta\rho_r/\bar{\rho}_r' $,
$\theta_x(\tau_{in})=0$. In both cases the background densities
corresponds to present values $\Omega_x=0.70$, $\Omega_c=0.2538$,
$\Omega_b=0.0462$, the initial curvature perturbation
$\zeta(\tau_{in})$ equal to $1\times 10^{-25}$ and comoving wave
number is $k = 7 \times 10^{-5}$ Mpc${}^{-1}$.}\label{fig:2}
\end{figure}

To verify the analytical conclusions we have modified the public
available CAMB code \cite{CAMB}. The initial adiabatic conditions
for all non-dark species are imposed the same as in the
non-interacting case. The initial values of dark sector
perturbations are taken in accordance with $
\delta\rho_c/\bar{\rho}_c'= \delta\rho_x/\bar{\rho}_x'=
\delta\rho_r/\bar{\rho}_r'$. In the presence $\gamma_x \neq 0$ the
cold dark mater rest frame and synchronous frame are not coincide
and it is not possible to adopt the CAMB conventions
$\theta_c\equiv 0$ consistently. In our numerical calculations the
residual synchronous gauge freedom was fixed by choosing
$\theta_c(\tau_{in}) =0$, where $\tau_{in}$ is the time moment of
the initial conditions setting. The results are shown in Figure
\ref{fig:2}.

\section{Special case.}
\label{sec:4}

The special case of (\ref{coupl_1}) is the interaction
\begin{equation}
\label{coupl_3}Q^\mu =\gamma \rho_c\rho_x (u^\mu_c-u^\mu_{x}).
\end{equation}
In the linear perturbation theory the velocity potential is a
first order quantity, and hence the spatial components of 4-vector
$Q^\mu$ are proportional to the product of the densities of both
dark species and relative velocity. In components
\begin{equation}
Q^0 =0,~~~~~~~Q^i =\frac{\gamma \rho_c\rho_x}{a} (v_{c}^{,i} -
v_{x}^{,i}).
\end{equation}

The background dynamics here is the same as at minimal coupling,
but the perturbation evolution is different. The coefficients of
second-order equation (\ref{de_pert}) for the large scale
perturbations are
\begin{equation}
A_x= 1-3w_x -\frac{2{\cal H}'}{{\cal H} ^2} + \frac{\gamma a
\bar{\rho}_c}{(1+w_x){\cal H}} - f,
\end{equation}
\begin{equation}
B_x= 3\left((1- w_x)\left(1 - \frac{{\cal H}'}{{\cal H} ^2}- f
\right)- \frac{w_xw_x'}{(1+w_x){\cal H} } +\frac{ 1-w_x }{1 +w_x}
\frac{\gamma a\bar{\rho}_c }{{\cal H}} \right) .
\end{equation}
Since now $\bar{\rho}_c =\rho_{c0}a^{-3}$, we have $|\gamma a
\bar{\rho}_c/{\cal H}|\gg 1$ in early Universe. Thus, at $ \gamma
<0$ the coefficients $A_x$, $B_x$ are both negative and there is a
rapid growth of non-adiabatic perturbations in the radiation
dominated era. At positive $\gamma$ the non-adiabatic growth of
long-wavelength perturbations do not occurs. These analytical
results are confirmed by numerical computations.

In the short-wave approximation the equations (\ref{basic_eq1}) -
(\ref{Einstein_pert}) can be combined into
\begin{eqnarray}
\label{eq_small1}\delta_x ''+ {\cal H}\left( 1-3w_x-\frac{w_x'
}{(1 + w_x){\cal H} }  +  \frac{\gamma a\rho_c}{(1 + w_x){\cal
H}}\right)\delta_x '+ k^2 \delta_x \nonumber \\
= {\cal H}^2 (1 + w_x)\left(\frac{3}{2} \left(\delta +
3\frac{\delta P}{\rho} \right) - \left(3 - \frac{\gamma a
\rho_c}{(1 + w_x){\cal H}}\right)\frac{\delta_c '}{{\cal H}} - 3
\frac{\theta_c}{{\cal H}} \right),
\end{eqnarray}
\begin{eqnarray}
\label{eq_small2}\delta_c '' +{\cal H}\left(1+ \frac{\gamma
a\bar{\rho}_x}{{\cal H}}
\right) \delta_c '  \nonumber\\
= {\cal H}^2\left(\frac{3}{2} \left(\delta + 3\frac{\delta P}{
\rho} \right) +\frac{\gamma a\bar{\rho}_x}{(1 + w_x){\cal
H}}\left(3(1-w_x) \delta_x +\frac{\delta_x'}{{\cal H}}
\right)\right).
\end{eqnarray}

At positive $\gamma$ all coefficients in the left hand side of
these equations are also positive, what excludes the presence of
small scale adiabatic instabilities. For instance, at early times
in the strong coupling regime with $\frac{a\gamma
\bar{\rho}_c}{{\cal H}}{\cal H}^2\gg k^2\gg {\cal H}^2$ the first
equation gives at leading order
\begin{equation}
\delta_x'=(1+w_x)\delta_c'.
\end{equation}

Equations (\ref{eq_small1}) and (\ref{eq_small2}) take the same
form as in the minimal coupling case under conditions
$\frac{a\gamma \bar{\rho}_c}{{\cal H}} \ll 1$ and
$\frac{a\gamma\bar{\rho}_x}{{\cal H}} \ll 1$ respectively. In
particular, when $\gamma\rightarrow 0$, the second one reduced to
the standard growth equation.

\section{Conclusions.} \label{sec:5}

We examined the covariant generalization of the coupling
(\ref{coupl_backgr2}). The evolution of perturbations is studied
paying particular attention to the most favored interaction rate
that is proportional to the product of dark matter and dark energy
densities. It is shown that the models of the form (\ref{coupl_2})
with $b=1$ and $w_x(\tau)<-1/3$ in radiation dominated era suffers
from early time instabilities due fast growth of large scale
non-adiabatic perturbations. Models with $w_x(\tau)>-1/3$ in
radiation dominated Universe are free from this defect. Also an
interesting coupling (\ref{coupl_3}) with positive $\gamma$ is
viable.


\begin{thebibliography}{}
\bibitem{Astier}
P. Astier et al., "The Supernova legacy survey: Measurement of
omega(m), omega(lambda) and W from the first year data set",
Astron. Astrophys., \textbf{447}, 31–48 (2006).

\bibitem{WMAP5}
E. Komatsu et al., "Five-Year Wilkinson Microwave Anisotropy Probe
(WMAP) Observations: Cosmological Interpretation", Astrophys. J.
Suppl. \textbf{180}, 330-376 (2009).

\bibitem{SDSS}
M. Tegmark et al., "Cosmological Constraints from the SDSS
Luminous Red Galaxies", Phys. Rev. \textbf{D74}, 123507 (2006).

\bibitem{Weinberg}
S. Weinberg,"The Cosmological Constant Problem", Rev. Mod. Phys.
\textbf{61}, 1-23 (1989).

\bibitem{Hoffman_rev} Mark B. Hoffman "Cosmological constraints
on a dark matter – dark energy interaction",
arXiv:astro-ph/0307350 .

\bibitem{Wetterich}
C. Wetterich , "The cosmon model for an asymptotically vanishing
time-dependent cosmological constant",  Astron. Astrophys.
\textbf{301}, 321-328  (1995).

\bibitem{Amendola}
L. Amendola,"Scaling solutions in general non-minimal coupling
theories", Phys. Rev. \textbf{D60},  043501  (1999).

\bibitem{Holden_Wands}
D.J. Holden and D. Wands,"Self-similar cosmological solutions with
a non-minimally coupled scalar field" ,Phys. Rev. \textbf{D61},
043506 (2000).

\bibitem{Farrar_Peebles}
Glennys R. Farrar, and P. J. E. Peebles "Interacting Dark Matter
and Dark Energy", Astrophys.J. \textbf{604}, 1-11 (2004).

\bibitem{Ziaeepour}
H. Ziaeepour, "Quintessence From The Decay of a Superheavy Dark
Matter", Phys. Rev. \textbf{D69}, 063512 (2004).

\bibitem{Koivisto}
T. Koivisto "Growth of perturbations in dark matter coupled with
quintessence", Phys. Rev. \textbf{D72} 043516 (2005).

\bibitem{08011105}
R. Bean , E. E. Flanagan , I. Laszlo  and M. Trodden,
"Constraining Interactions in Cosmology's Dark Sector", Phys. Rev.
\textbf{D78}, 123514 (2008).

\bibitem{CMU}
G. Caldera-Cabral, R. Maartens, L. A. Urena-Lopez, "Dynamics of
interacting dark energy", Phys. Rev. \textbf{D79}, 063518 (2009).

\bibitem{Pavon2003}
L.P. Chimento, A.S. Jakubi, D. Pavon and W. Zimdahl, "Interacting
quintessence solution to the coincidence problem", Phys. Rev.
\textbf{D67},  083513  (2003).

\bibitem{QCJRW}
M. Quartin, M.O. Calvao, S.E. Joras,  R.R. Reis and I. Waga, "Dark
Interactions and Cosmological Fine-Tuning", JCAP \textbf{0805},
007 (2008).

\bibitem{VMM_0804}
J. Valiviita, E. Majerotto  and R. Maartens, "Large-scale
instability in interacting dark energy and dark matter fluids",
JCAP \textbf{0807}, 020 (2008).

\bibitem{08073471}
Jian-Hua He, B. Wang, E. Abdalla, "Stability of the curvature
perturbation in dark sectors' mutual interacting models", Phys.
Lett. \textbf{B671}, 139-145 (2009).

\bibitem{09011611}
M.B. Gavela, D. Hernandez,  L. Lopez Honorez, O. Mena and S.
Rigolin, "Dark coupling", JCAP \textbf{0907}, 034 (2009).

\bibitem{09013272}
B. M. Jackson, A. Taylor, A. Berera, "On the large-scale
instability in interacting dark energy and dark matter fluids",
Phys. Rev. \textbf{D79}, 043526 (2009).

\bibitem{09050492}
G. Caldera-Cabral, R. Maartens, B. M. Schaefer, "The Growth of
Structure in Interacting Dark Energy Models",  JCAP \textbf{0907},
027 (2009).

\bibitem{VMM_0907}
J. Valiviita, E. Majerotto  and R. Maartens, "Observational
constraints on an interacting dark energy model", Mon. Not. Roy.
Astron. Soc. \textbf{402}, 2355-2368 (2010).

\bibitem{0212518}
G. Mangano, G. Miele and V. Pettorino, "Coupled quintessence and
the coincidence problem",  Mod. Phys. Lett. \textbf{A18}, 831
(2003).

\bibitem{coincidence_original}
P. J. Steinhardt, "Cosmological Challenges for the  21st Century",
in "Critical Problems in Physics", edited by V. L. Fitch, D. R.
Marlow, and M. A. E. Dementi, p.123, Princeton  U. Press,
Princeton (1997)

\bibitem{coincidence} I. Zlatev, L. Wang and P. J.
Steinhardt "Quintessence, Cosmic Coincidence, and the Cosmological
Constant",  Phys. Rev. Lett. \textbf{82}, 896-899 (1999).

\bibitem{09011215}
Yin-Zhe Ma,  Y. Gong,  and X. Chen, "Couplings between holographic
dark energy and dark matter", arXiv:0901.1215 [astro-ph].

\bibitem{parametrization}
M. Chevallier and D. Polarski, "Accelerating  Universes with
Scaling Dark Matter", Int. J.Mod. Phys. \textbf{D10}, 213 (2001).

\bibitem{Kodama_Sasaki}
H. Kodama and M. Sasaki , "Cosmological Perturbation Theory",
Prog. Theor. Phys. Suppl. \textbf{78}, 1 (1984)

\bibitem{Ma_Bertschinger}
Chung-Pei Ma , E. Bertschinger, "Cosmological Perturbation Theory
in the Synchronous and Conformal Newtonian Gauges", Astrophys. J.
\textbf{455}, 7-25 (1995).

\bibitem{Gordon_Hu}
C. Gordon and W. Hu, "A Low CMB Quadrupole from Dark Energy
Isocurvature Perturbations", Phys.Rev. \textbf{D70} 083003 (2004).

\bibitem{GWBM}
C. Gordon, D. Wands, B. A. Bassett and R. Maartens, "Adiabatic and
entropy perturbations from inflation", Phys. Rev. \textbf{D63}
023506 (2001).

\bibitem{CAMB}
A. Lewis, A. Challinor, and A. Lasenby, "Efficient computation of
CMB anisotropies in closed FRW models", Astrophys. J.,
\textbf{538}, 473-476 (2000).



\end{thebibliography}
\end{document}